\documentclass[a4paper]{article}

\usepackage{INTERSPEECH2020}

\usepackage{times}
\usepackage{soul}
\usepackage{url}
\usepackage[hidelinks]{hyperref}
\usepackage[utf8]{inputenc}
\usepackage{graphicx}
\usepackage{amsmath}
\usepackage{amsthm}
\usepackage{booktabs}
\usepackage{algorithm}
\usepackage{algorithmic}
\usepackage{subcaption}
\title{GAZEV: GAN-Based Zero-Shot Voice Conversion over Non-parallel Speech Corpus}
\name{Zining Zhang$^{1,2}$, Bingsheng He$^2$, Zhenjie Zhang$^1$}
\address{
  $^1$Singapore R\&D, Yitu Technology\\
  $^2$School of Computing, National University of Singapore}
\email{zining.zhang@yitu-inc.com, hebs@comp.nus.edu.sg, zhenjie.zhang@yitu-inc.com}

\begin{document}

\maketitle
\begin{abstract}
Non-parallel many-to-many voice conversion is recently attracting huge research efforts in the speech processing community.
A voice conversion system transforms an utterance of a source speaker to another utterance of a target speaker by keeping the content in the original utterance and replacing by the vocal features from the target speaker.
Existing solutions, e.g., StarGAN-VC2, present promising results, \emph{only} when speech corpus of the engaged speakers is available during model training. 
AUTOVC is able to perform voice conversion on unseen speakers, but it needs an external pretrained speaker verification model.
In this paper, we present our new GAN-based zero-shot voice conversion solution, called GAZEV, which targets to support unseen speakers on both source and target utterances. Our key technical contribution is the adoption of speaker embedding loss on top of the GAN framework, as well as adaptive instance normalization strategy, in order to address the limitations of speaker identity transfer in existing solutions. Our empirical evaluations demonstrate significant performance improvement on output speech quality, and comparable speaker similarity to AUTOVC.
\end{abstract}
\noindent\textbf{Index Terms}: voice conversion, generative adversarial network, zero-shot, non-parallel

\section{Introduction}\label{sec:intro}

The emergence and development of generative adversarial network, or GAN in short, has enabled efficient and effective style transfer over image and audio domains. 
Motivated by the huge success of general GAN technology, GAN-based voice conversion has recently attracted extensive research efforts in the speech processing community. In voice conversion, the speaker identity is regarded as a special type of style, such that the conversion becomes a transfer from one speaker identity to another, as a special case of style transfer.
A voice conversion system transforms an utterance of a source speaker to the utterance of a target speaker by keeping the content in the original utterance and replacing by the vocal features with the target speaker.
The adoption of CycleGAN and StarGAN in speech synthesis models has brought us the state-of-the-art solutions of voice conversion, e.g., CycleGAN-VC \cite{kaneko2018cyclegan,kaneko2019cyclegan} and StarGAN-VC \cite{kameoka2018stargan, kaneko2019stargan}. 

Although these approaches have demonstrated very promising results, they work only when both the source speaker and the target speaker are present in the training dataset.
This requirement limits the usage of voice conversion in real applications.
In this paper, we discuss how to circumvent the limitation and achieve the objective of zero-shot voice conversion on non-parallel speech corpus.

Our new approach, called GAZEV, is based on StarGAN-VC.
In GAZEV, we extend the StarGAN-VC model architecture to handle zero-shot many-to-many conversion, in the sense that the model is able to convert between speeches from arbitrary speakers, regardless of the speaker's presence in the speech corpus for model training.

GAZEV introduces two new methods into the structure of StarGAN-VC to support unseen speakers in many-to-many voice conversion. Firstly, we design and insert a customized adaptive instance normalization operator into StarGAN-VC. Such an operator allows the model to better capture the impact of speaker identity in the generative process of voice conversion. Secondly, we add an additional speaker embedding loss into the model training.
This speaker embedding loss ensures that the embedding of the converted audio is close to the target speaker embedding, in addition, it also helps avoid the conversion output to collapse to similar voices under different speaker embeddings.

The combination of the new methods has greatly enhanced the performance of voice conversion, especially in the most challenging cases with unseen speakers. To summarize, we list the core contributions of the paper as follows:

\begin{enumerate}
    \item We present a new approach based on StarGAN-VC for zero-shot many-to-many voice conversion problems.
    \item We design a customized adaptive instance normalization operator for voice conversion task.
    \item We revise the loss function to better address the demand of unseen speakers during inference.
    \item We conduct empirical studies of our proposed approach and compare it against state-of-the-art solutions on zero-shot voice conversion.
\end{enumerate}

The rest of the paper is organized as follows. Section \ref{sec:model} presents the model structure and details of the training and inference algorithms. Section \ref{sec:experiment} reports the empirical results of our approach on a dataset composed of 109 different speakers. Section \ref{sec:related} reviews the existing studies on related topics and finally, Section \ref{sec:conclusion} concludes the paper and discusses the future research directions.

\section{Model}\label{sec:model}

In this section, we present the model architecture of GAZEV, in comparison with StarGAN-VC. Assume we have $n$ speakers in the dataset with ids from $1$ to $n$, i.e., $\mathbb{N}=\{1,2,\ldots,n\}$, and a speech audio domain $\mathbb{A}$, each audio of length $T$ in $\mathbb{A}$ is a sequence $\boldsymbol{x}(t)$ with $t=1,2,\ldots,T$. When the context is clear, we abuse $\boldsymbol{x}$ and $\boldsymbol{y}$ to denote audio sequence $\boldsymbol{x}(t)$ and $\boldsymbol{y}(t)$. Given an input audio sequence $\boldsymbol{x}$ from speaker id $c_x\in\mathbb{N}$ and a target speaker id $c_y\in\mathbb{N}$, the objective of voice conversion is to generate a high-quality audio sequence $\boldsymbol{\hat{y}}$ capturing the linguistic contents from $\boldsymbol{x}$ and vocal features of speaker $c_y$. Moreover, we use $u_x$ and $u_y$ to denote the gender of the speaker with utterance $\boldsymbol{x}$ and $\boldsymbol{y}$ respectively.

\subsection{Architecture of StarGAN-VC}\label{subsec:starganvc}

StarGAN-VC is a generalization of CycleGAN-VC, by employing the successful framework of StarGAN in computer vision \cite{choi2018stargan} instead of the framework of CycleGAN \cite{zhu2017unpaired}.

\begin{figure}[t]
    \centering
        \includegraphics[width=.94\linewidth]{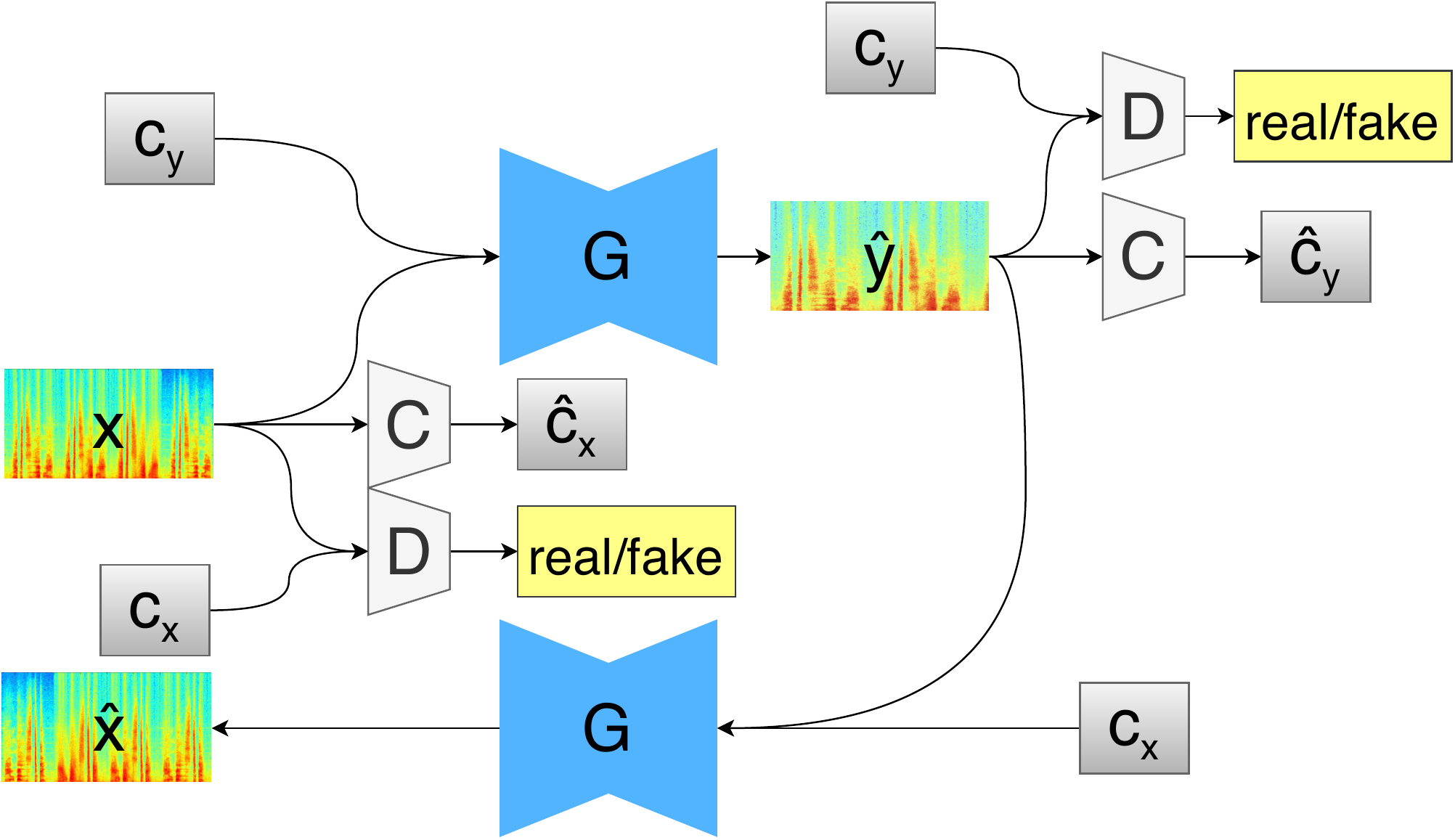}
        \caption{The architecture of StarGAN-VC: It includes an autoencoder model $G$, a classification model $C$ and a discriminator model $D$.}
        \label{fig:starganvc}
        \vspace{-10pt}
\end{figure}

In Figure \ref{fig:starganvc}, we present the model architecture of StarGAN-VC. 
Given the source utterance $\boldsymbol{x}$ and the target speaker id $c_y$, the model $G$ generates the estimated utterance $\boldsymbol{\hat{y}}$ based on the vocal features of speaker $c_y$, i.e., $\boldsymbol{\hat{y}}=G(\boldsymbol{x},c_y)$. 
In order to train the model $G$, StarGAN-VC employs and trains a speaker classifier $C$ and a discriminator $D$. 
Given an arbitrary speech audio clip, e.g., $\boldsymbol{\hat{y}}$, the classifier $C$ attempts to retrieve the id of the corresponding speaker. 
The discriminator $D$ takes a speaker id, e.g., $c_y$, and an audio clip, e.g., $\boldsymbol{\hat{y}}$, as inputs, and tries to detect if the audio clip is real human voice or synthesized voice. 
Following the strategy of CycleGAN-VC and StarGAN-VC, these approaches expect the model to generate highly natural speech audio, such that discriminator is unable to tell the difference between real voice $\boldsymbol{x}$ and fake voice $\boldsymbol{\hat{y}}$. 
Moreover, the inclusion of classifier $C$'s performance on id retrieval in the loss function encourages the model to generate speech similar to the utterance made by the target speaker.

Generally speaking, the loss function of StarGAN-VC consists of three parts, namely \emph{adversarial loss}, \emph{classification loss}, \emph{cycle consistency loss}, and \emph{identity mapping loss}. 
Specifically, the adversarial loss indicates how the generative model $G$ could confuse the discriminator $D$, with the expectations calculated with variables over $c$ and $\boldsymbol{x}$, 

\begin{equation}
    \begin{split}
        \mathcal{L}_{adv} & := -\mathbb{E}_{c_x\sim p(c), \boldsymbol{x}\sim p(\boldsymbol{x}|c_x)}[\log(D(\boldsymbol{x}, c_x))] \\
        & \;\;\;\; -\mathbb{E}_{\boldsymbol{x}\sim p(\boldsymbol{x}), c_y\sim p(c)}[\log(1-D(G(\boldsymbol{x}, c_y),c_y))]
    \end{split}
\end{equation}

The classification loss is the expectation of classifying the utterance to a wrong speaker, which is applied to both original utterance $\boldsymbol{x}$ and generated utterance $\boldsymbol{\hat{y}}$.

\begin{equation}
    \begin{split}
        \mathcal{L}^C_{cls} & := -\mathbb{E}_{c_x\sim p(c), \boldsymbol{x}\sim p(\boldsymbol{x}|c_x)}[\log(Pr(c_x=C(\boldsymbol{x})))] \\
    \end{split}
\end{equation}
\begin{equation}
    \begin{split}
        \mathcal{L}^G_{cls} & := -\mathbb{E}_{\boldsymbol{x}\sim p(\boldsymbol{x}), c_y\sim p(c)}[\log(Pr(c_y=C(G(\boldsymbol{x}, c_y))))]
    \end{split}
\end{equation}

Finally, the cycle consistency loss, which expects the model to work well when converting the utterance back to its original speaker; and identity mapping loss, the reconstruction loss of converting the voice to the same speaker.

\begin{equation}
    \begin{split}
        \mathcal{L}_{cyc} & := -\mathbb{E}_{c_x \sim p(c), \boldsymbol{x}\sim p(\boldsymbol{x}|c_x), c_y\sim p(c)}[\Vert G(G(\boldsymbol{x}, c_y), c_x)  - \boldsymbol{x}\Vert]
    \end{split}
\end{equation}
\begin{equation}
    \begin{split}
        \mathcal{L}_{id} & := -\mathbb{E}_{c_x \sim p(c), \boldsymbol{x}\sim p(\boldsymbol{x}|c_x)}[\Vert G(\boldsymbol{x}, c_x)  - \boldsymbol{x}\Vert]
    \end{split}
\end{equation}

\subsection{Architecture of GAZEV}

\begin{figure}[t]
    \centering
        \includegraphics[width=.94\linewidth]{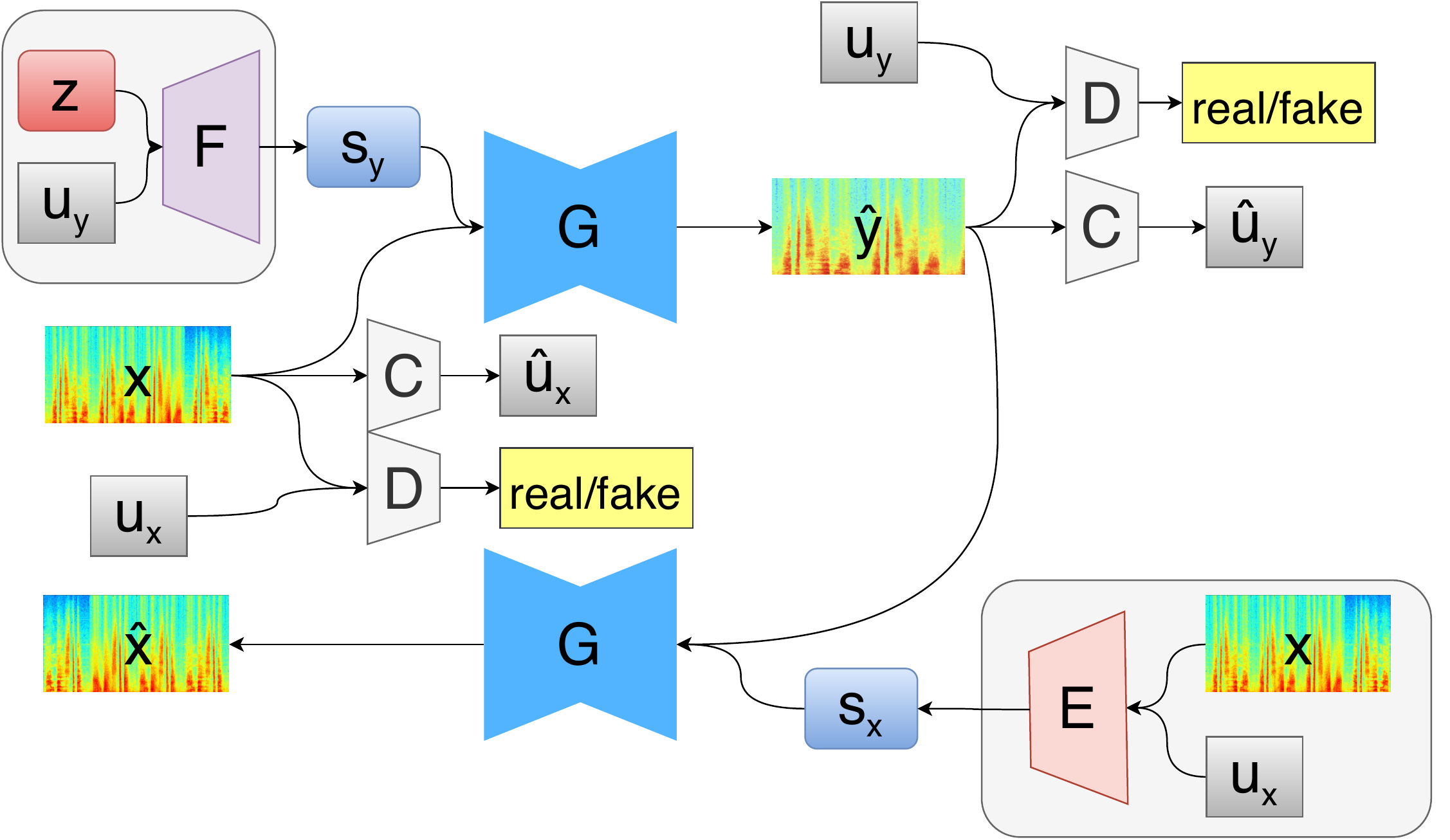}
        \caption{Model Architecture of GAZEV: Compared to StarGAN-VC, we use speaker embedding instead of speaker id. Speaker embedding is calculated by $F$ based on gender $u_y$ of speaker $c_y$ and a Gaussian prior $z$, or by $E$ based on its original utterance $x$ and gender information $u_x$ of speaker $c_x$.}
        \label{fig:gazev}
        \vspace{-10pt}
\end{figure}

While StarGAN-VC performs non-parallel many-to-many voice conversion, the source and target speaker must appear in the training data.
To enable zero-shot voice conversion possible, we make a few revisions to the model architecture in GAZEV, as illustrated in Figure \ref{fig:gazev}. We introduce the speaker embedding vectors $\boldsymbol{s}_x$ and $\boldsymbol{s}_y$ into the model, corresponding to utterance $\boldsymbol{x}$ and $\boldsymbol{y}$ respectively. The speaker embedding is expected to provide more information than the speaker id, which is the key to our extension to zero-shot voice conversion. 

There are two different methods of generating speaker embedding. 
A model $F(\boldsymbol{z},u_y)$ generates the embedding vector $\boldsymbol{s}_y$ by taking gender of $\boldsymbol{y}$ and a randomized Gaussian prior $\boldsymbol{z}$ which follows a unit Gaussian distribution. 
Another model $E(\boldsymbol{x},u_x)$ is deployed to encode an input utterance $\boldsymbol{x}$ and its gender to the speaker embedding.

Another important revision to the original StarGAN-VC architecture is the extensive replacement of speaker id $c_y$ with gender attribute $u_y$.
The prediction target of classifier $C$, for example, is now the gender of the speaker in the utterance.
This greatly simplifies the classifier, consequently making it easier to reduce the classification loss and contributing more to the optimization of the generative model $G$.

In order to exploit the rich information in speaker embedding, we further introduce the strategy of AdaIn into the generative model $G$ \cite{ulyanov2016instance}.
The key idea of AdaIn is a special case of instance normalization, with scaling and bias variables controlled by the style embedding. In GAZEV, the style of speech is actually the speaker embedding. This leads to the insertion of the following operator into the generative model $G$, replacing the standard normalization operator after the convolution operations:

\begin{equation}
    \mbox{AdaIN}(\boldsymbol{x}, \boldsymbol{s}) = f({\boldsymbol{s}}) \Big(\frac{\boldsymbol{x}-\mu(\boldsymbol{x})}{\sigma(\boldsymbol{x})}\Big)+g(\boldsymbol{s}),
\end{equation}

in which $\boldsymbol{x}$ is an input to the operator, $\boldsymbol{s}$ is the input speaker embedding, $f(\boldsymbol{s})$ and $g(\boldsymbol{s})$ are two affine transformations applied on $\boldsymbol{s}$ to generate style-dependent scaling and bias factor.

By adding this AdaIN architecture to our generator, the new generator can be defined as $G(\boldsymbol{x}, \boldsymbol{s_y})$, which takes an embedding vector $\boldsymbol{s_y}$ as inputs instead of the speaker id $c_y$. 
The discriminator $D(\boldsymbol{x}, u)$ is defined to predict whether the input audio is real or fake under the specific gender, male or female.
And the classifier $C(\boldsymbol{x})$ is used to infer whether the audio is from a voice of male or female.

Based on the definitions above, we redefine the loss function to reflect the change of model architecture. The loss function in the training of the model consists of five parts, including \emph{adversarial loss}, \emph{classification loss}, \emph{cycle consistency loss}, \emph{identity mapping loss}, and \emph{speaker embedding loss}. The first four types of losses are formulated as follows, with $\boldsymbol{z} \sim \mathcal{N}(\boldsymbol{0}, \boldsymbol{I})$:
\begin{equation}
    \begin{split}
        \mathcal{L}_{adv} := &-\mathbb{E}_{u_x \sim p(u), \boldsymbol{x} \sim p(\boldsymbol{x}|u_x) }[\log(D(\boldsymbol{x}, u_x)] \\
        &-\mathbb{E}_{\boldsymbol{x} \sim p(\boldsymbol{x}), u_y \sim p(u) }[\log(1- \\
        &\;\;\;\;D(G(\boldsymbol{x}, F(\boldsymbol{z}, u_y)),u_y))] \\
    \end{split}
\end{equation}
\begin{equation}
    \begin{split}
        \mathcal{L}_{cls}^C := &-\mathbb{E}_{u_x \sim p(u), \boldsymbol{x} \sim p(\boldsymbol{x}|u_x)}[\log(\Pr(u_x=C(\boldsymbol{x}))] \\
    \end{split}
\end{equation}
\begin{equation}
    \begin{split}
        \mathcal{L}_{cls}^G := &-\mathbb{E}_{\boldsymbol{x} \sim p(\boldsymbol{x}), u_y \sim p(u) }[\log(\Pr(u_y=C(G(\boldsymbol{x}, F(\boldsymbol{z}, u_y))))]
    \end{split}
\end{equation}
\begin{equation}
\begin{split}
    \mathcal{L}_{cyc} := &-\mathbb{E}_{u_x \sim p(u), \boldsymbol{x} \sim p(\boldsymbol{x}|u_x), u_y \sim p(u)}[ \\
    & \Vert G(G(\boldsymbol{x}, F(\boldsymbol{z}, u_y)), E(\boldsymbol{x}, u_x))  - \boldsymbol{x}\Vert]
\end{split}
\end{equation}
\begin{equation}
    \mathcal{L}_{id} := -\mathbb{E}_{u_x \sim p(u), \boldsymbol{x} \sim p(\boldsymbol{x}|u_x)}[\Vert G(\boldsymbol{x}, E(\boldsymbol{x}, u_x)) - \boldsymbol{x}\Vert]
\end{equation}

Speaker embedding loss includes the error on the embedding when encoder $E$ calculates the speaker embedding over the converted speech audio and compares it against the true speaker embedding of the target speaker. We also try to maximize the divergence of output speeches under different speaker embeddings, to avoid the voices to collapse. This leads to the following definition:

\begin{equation}
\begin{split}
        \mathcal{L}_{spk} := &-\mathbb{E}_{\boldsymbol{x} \sim p(\boldsymbol{x}), u_y \sim p(u)}[\\ 
        &\Vert E(G(\boldsymbol{x}, F(\boldsymbol{z}, u_y)), u_y) - F(\boldsymbol{z}, u_y) \Vert] \\
        &+\mathbb{E}_{\boldsymbol{x} \sim p(\boldsymbol{x}), u_y \sim p(u)}[\\
        &\Vert G(\boldsymbol{x}, F(\boldsymbol{z_1}, u_y)) - G(\boldsymbol{x}, F(\boldsymbol{z_2}, u_y)) \Vert]
\end{split}
\end{equation}

\section{Experiments}\label{sec:experiment}

\subsection{Dataset}\label{subsec:dataset}
In our experiments, we train GAZEV and our baseline models with English Multi-speaker Corpus for CSTR Voice Cloning Toolkit (VCTK) \cite{veaux2016superseded}, which contains 109 native English speakers. In VCTK, there are around 400 utterances for each speaker.
80 speakers are used in our training dataset. 
The MCC(mel cepstral coefficients)s are extracted using WORLD vocoder \cite{morise2016world}, with dimension 36, and each record is segmented with length 256 frames.
In the testing dataset, we include 10 seen speakers and 10 unseen speakers. For both seen and unseen speakers, the genders are evenly distributed. We run our testings to cover all different cases, between speakers of different genders, as well as between seen and unseen speakers.

\subsection{Model Setup}\label{subsec:modelsetup}
The model is composed of five trainable components: the \textit{Generator} (\textit{G}), the \textit{Discriminator} (\textit{D}), the \textit{Classifier} (\textit{C}), the \textit{Speaker Embedding Generator} (\textit{F}), and the \textit{Speaker Encoder}(\textit{E}).

The \textit{Generator} consists of three parts,  all of which use convolution neural networks (CNN):  2 down-sampling blocks, with initial channel number 64, and kernel size (4,8), down-sampling with a factor of 2;
conversion blocks, with first 3 residual blocks using instance normalization, and next 3 residual blocks using AdaIN with affine transformations, the channel size is 256, and kernel size is 3;
2 up-sampling blocks, with initial channel size 256, and kernel size (4,4), up-sampling with a factor of 2.

For the \textit{Discriminator}, we employ the discriminator from PatchGAN \cite{shi2016real, li2016precomputed, isola2017image}. Instead of applying the fake/real classification over the complete audio clip, PatchGAN attempts to classify over patches or segments of audios. This strategy increases the difficulty of classifier and turns out to be effective in improving the performance of standard discriminator in voice transfer algorithms \cite{kaneko2019cyclegan}.
The \textit{Discriminator} is composed of 5 down-sampling blocks with initial channel size 64, kernel size 4, and down-sampling with a factor of 2. The \textit{Classifier} shares the layers except the output layer with the \textit{Discriminator}.

\textit{Speaker Embedding Generator} is simply a 5 layers MLP with dimension 512, and ReLU activation function. \textit{Speaker Encoder} is composed of multiple pre-activation residual blocks as defined in \cite{he2016identity}. There are 5 such residual blocks, each with channel size 32 and kernel size 3.
For the random sample $\boldsymbol{z}$ and speaker embedding $\boldsymbol{s}$, the dimensions are set to 16 and 64 respectively.

When training GAZEV, the weight over the \emph{discriminator loss} is 1, while the weights for all other losses are 10.
Each batch contains 32 samples in the training, and the learning rate is 0.0001. We test the performance of GAZEV after training 1,000,000 steps with Adam optimizer.

\subsection{Evaluation Criterion}\label{subsec:evaluationcriterion}

We employ AUTO-VC as our baseline approach in the experiments, because it is the only zero-shot voice conversion algorithm available for testing. AUTO-VC uses a simple auto-encoder architecture, with an external speaker verification model pre-trained following GE2E \cite{wan2018generalized}.
Ideally, the encoder in AUTO-VC encodes the content information of the input audio. The resulting content together with the target speaker embedding, calculated using the pre-trained model, are fed into the decoder to generate the output audio clip with identical linguistic content as well as the target speaker's voice.
To build the baseline model, we use the implementation from the authors of AUTO-VC. In testing, we take the model after 2,000,000 training steps. To make a fair comparison, the model of AUTO-VC is trained using the same 80 speakers' speech corpus as GAZEV does.

MOS test is used for evaluating both naturalness and similarity of the results.
The MOS score is from 1 to 5, with 0.5 increments. Therefore, there are 9 options for the evaluations.
Regarding naturalness MOS, human annotators are given with converted audios. The annotator needs to provide a MOS score to the quality of the speech audio, based on human's perception. A higher score means better speech naturalness.
Regarding similarity MOS, in addition to the sample audio, a reference audio of the target speaker's voice is also provided. Similar to naturalness MOS, the annotator compares the reference audio and the target audio before providing a score from 1 to 5.

\subsection{Performance Evaluation}\label{subsec:performance}
As stated in Section~\ref{subsec:dataset}, the test dataset consists of 10 seen speakers and 10 unseen speakers. To compose the 4 different configurations of voice conversion, there are 400 converted voices in the experiment result.
The result is further divided into 4 categories based on the genders of the speakers on both source and target sides, i.e., female-to-female (F2F), female-to-male (F2M), male-to-female (M2F) and male-to-male (M2M).
Moreover, the result is also divided into 4 categories based on the types of conversion between seen and unseen speakers, i.e., seen-to-seen (S2S), seen-to-unseen (S2U), unseen-to-seen (U2S) and unseen-to-unseen (U2U).

\begin{figure}[t]
    \centering
        \includegraphics[width=1.1\linewidth]{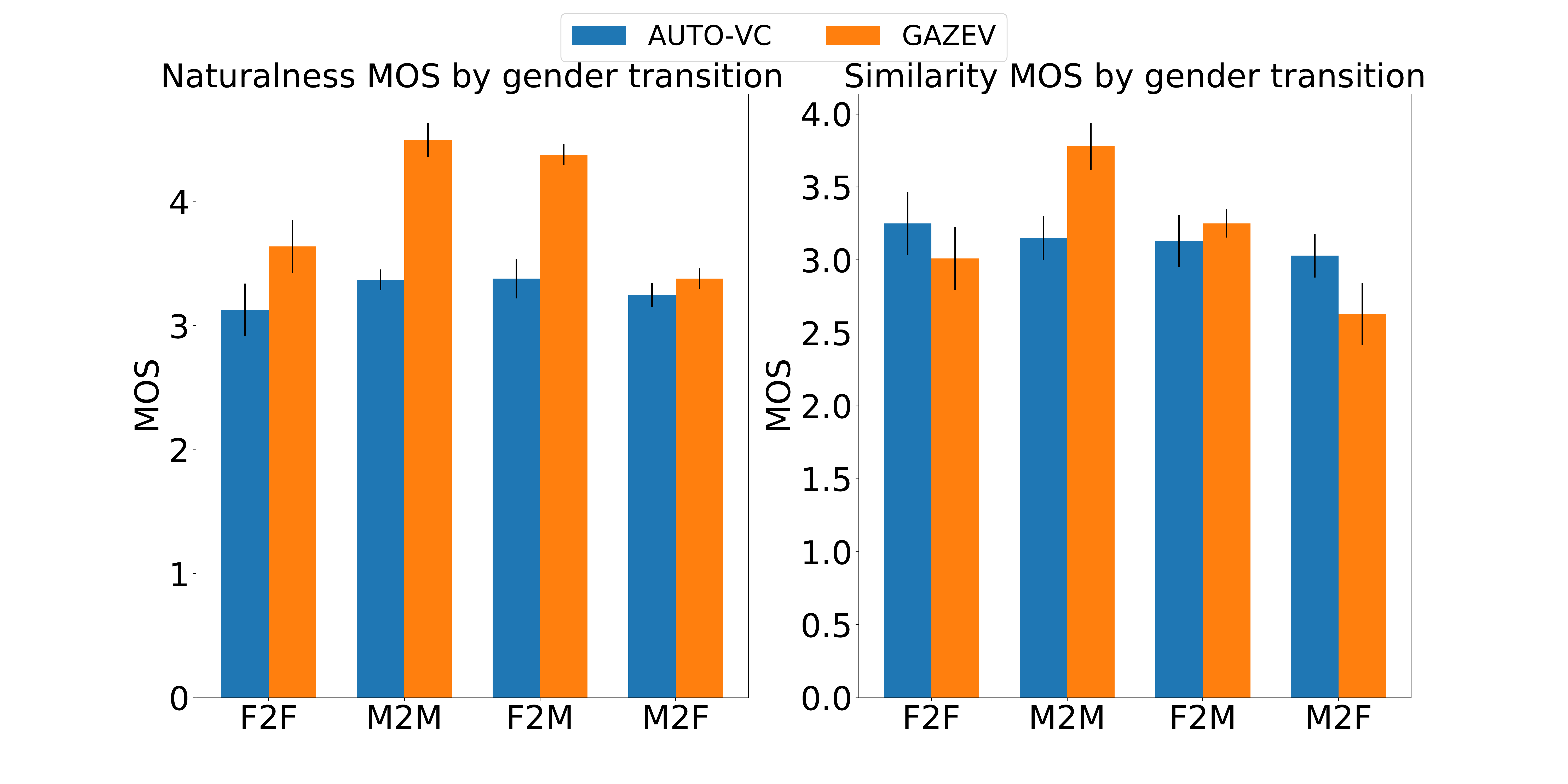}
        \vspace{-5pt}
        \caption{The naturalness and similarity MOS scores for AUTO-VC and GAZEV on gender conversion}
        \label{subfig:mosgen}
        \vspace{-10pt}
\end{figure}
\begin{figure}[t]
    \centering
        \includegraphics[width=1.1\linewidth]{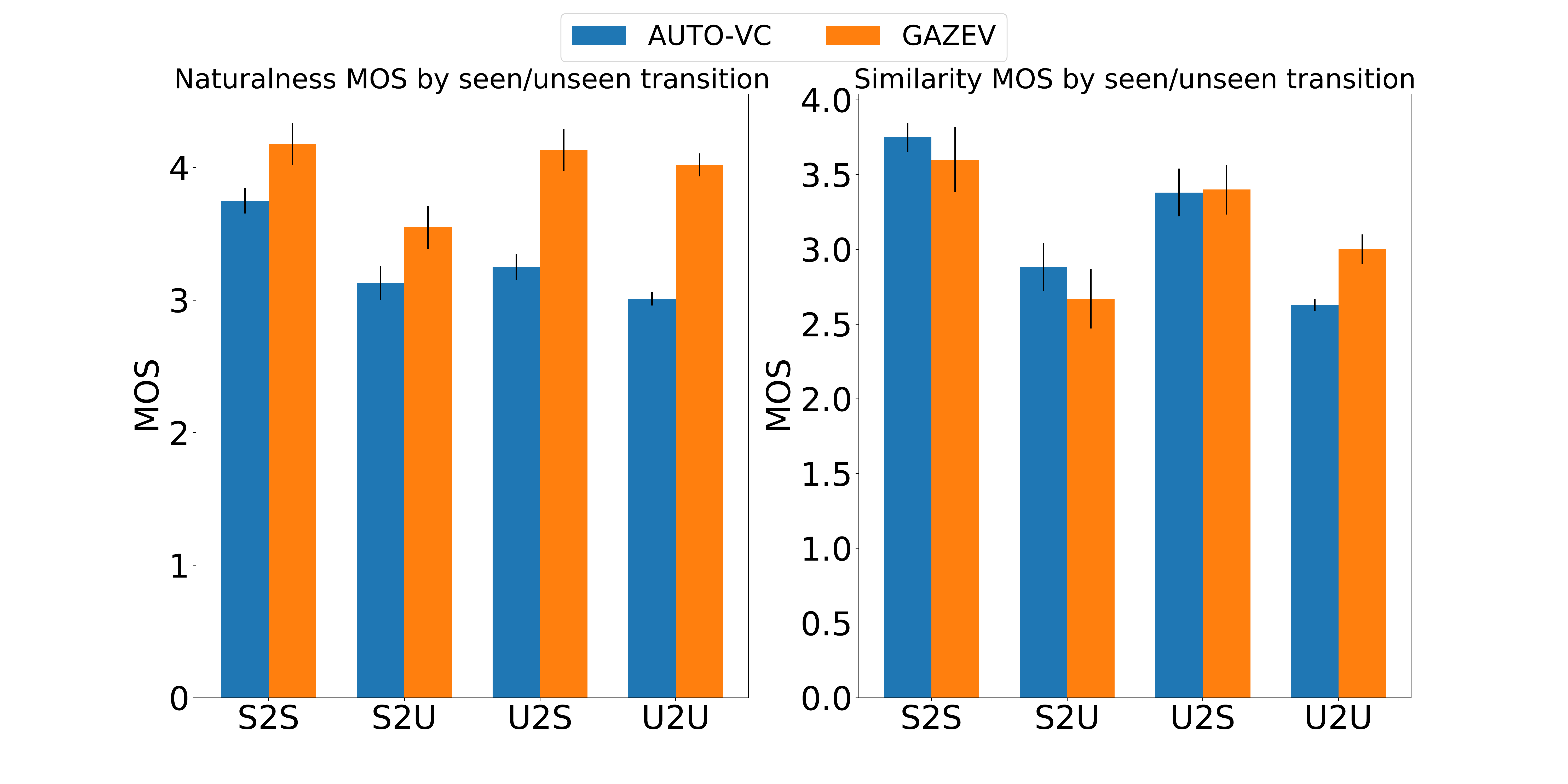}
        \vspace{-5pt}
        \caption{The naturalness and similarity MOS scores for AUTO-VC and GAZEV on seen/unseen conversion}
        \label{subfig:mossu}
        \vspace{-10pt}
\end{figure}
The test results\footnote{Demo samples are available at \url{https://speech-ai.github.io/gazev/}} are summarized in Figure~\ref{subfig:mosgen} and Figure~\ref{subfig:mossu}.
In the figures, GAZEV significantly outperforms AUTO-VC on speech naturalness. Note that the naturalness of the converted audios is above 4.0 in MOS results, even in the category of unseen-to-unseen (U2U).
It is a huge improvement over the only zero-shot voice conversion approach AUTO-VC in the literature.
GAZEV also achieves similar performance on similarity MOS, although AUTO-VC is equipped with a well pre-trained speaker encoder module. Note that the speaker encoder is trained on a much larger dataset with 3,549 speakers.
On the contrary, GAZEV learns a speaker embedding space only from these 80 speakers from the training data for voice conversion. We believe the performance of GAZEV could be even better, if a larger speech corpus is used in the training for our speaker embedding modules, i.e., $E$ and $F$ as in Figure \ref{fig:gazev}.

\section{Related Work}\label{sec:related}
The earlier voice conversion methods commonly use Gaussian mixture models(GMM) \cite{stylianou1998continuous, toda2007voice, helander2010voice}, restricted Boltzmann machine \cite{chen2014voice, nakashika2014voice}, feed-forward neural networks \cite{desai2010spectral, mohammadi2014voice,oyamada2017non}, recursive neural networks (RNN) \cite{nakashika2014high, sun2015voice}, and convolution neural networks (CNN) \cite{kaneko2017sequence}. 
All these methods need parallel training data, such that the training pairs are text aligned. In order to get desirable results, the training pairs also need to be time-aligned. The text alignment and time alignment restrictions make the training data extremely hard to collect.
To relax the limitation above, voice conversion models on non-parallel corpus are proposed.
CycleGAN-VC is a GAN based model performing voice conversion between two speakers without parallel training data. It is based on the CycleGAN model widely used in the image-to-image style transfer.
The cycle consistency loss in CycleGAN-VC is the key to the success of the model even without a parallel corpus for training.
However, CycleGAN-VC is designed for voice conversion between two speakers only, namely one-to-one voice conversion. 
To each pair of speakers, a CycleGAN-VC model must be independently trained.
StarGAN-VC is the first model To support many-to-many voice conversion on a non-parallel corpus. During training, the conversion may happen between any two speakers in the dataset, and the same loss in CycleGAN-VC is deployed.
Moreover, there is an additional classification loss introduced in StarGAN-VC, which encourages the model to generate speech audio similar to the target speaker's voice. 
There are multiple variants of CycleGAN-VC in the literature. In \cite{chou2018multi}, Chou et al. propose a two-stage training scheme on top of CycleGAN. In the first stage, the model introduces a speaker classifier, and uses a reconstruction loss and a classification loss. In the second stage, it introduces another classifier and uses GAN loss and a classification loss. 
CC-GAN \cite{lee2020many} is another variant of CycleGAN-VC, which uses additional conditional inputs of speaker labels to achieve many-to-many voice conversion.
However, StarGAN-VC performs better and is even simpler.

In the computer vision domain, StarGAN v2 \cite{choi2019stargan} is proposed as an enhanced version of StarGAN. It adopts similar tricks as in GAZEV, including the AdaIN operators and the style embedding with different styles.

On another research direction, variational auto-encoders (VAE) is often used in speech synthesis and voice conversion, such as 
vanilla VAE-VC \cite{hsu2016voice}; 
CDVAE-VC \cite{huang2018voice} utilizes cross-domain VAE on two different representations of the audios; 
ACVAE-VC \cite{kameoka2018acvae} uses an auxiliary classifier for the generator to generate voices to the correct speakers as StarGAN-VC does;
VAW-GAN \cite{hsu2017voice} adds a discriminator after the audio output and also adds the WGAN \cite{arjovsky2017wasserstein} loss.
However, the naturalness of the output speech by VAE-based approaches is way worse than GAN-based approaches. 
AUTO-VC is the latest attempt with a simple auto-encoder architecture. It is able to perform zero-shot voice conversion, since it uses a pretrained speaker encoder model providing the speaker embedding for an unseen target voice.

\section{Conclusion and Future Work}\label{sec:conclusion}

In this paper, we propose, GAZEV, a zero-shot voice conversion approach by extending the idea of StarGAN to support unseen speakers in inference. It beats the state-of-the-art solution, AUTO-VC, by a huge margin on speech naturalness, while achieving almost identical speaker similarity as AUTO-VC. We believe the speaker embedding is the key to the success of GAZEV. It can be further improved by adopting a better embedding module with more speech corpus data. The replacement of speaker id with speaker gender implies that a simpler classifier is beneficial to the generative model. However, it remains unclear how to find the best balance between the classifier complexity and generative model complexity.

\bibliographystyle{IEEEtran}

\bibliography{mybib}

\end{document}